\documentclass[pra,superscriptaddress,twocolumn]{revtex4-2}
\usepackage[utf8]{inputenc}
\usepackage{adjustbox}
\usepackage{amsfonts}
\usepackage{amsmath}
\usepackage{amsthm}
\usepackage{amssymb}
\usepackage{mathtools}
\usepackage{graphicx}
\usepackage[svgnames]{xcolor}
\usepackage{hyperref}

\newcommand{\mr}[1]{\mathrm{#1}}
\newcommand{\mb}[1]{\mathbf{#1}}
\newcommand{\hc}{\mathrm{H. c.}}
\newcommand{\vac}{\mathrm{vac}}
\newcommand{\sinc}{\mathrm{sinc}}
\newcommand{\veps}{\varepsilon}

\DeclarePairedDelimiter{\ket}{\lvert}{\rangle}

\DeclarePairedDelimiter{\braket}{\langle}{\rangle}
\DeclarePairedDelimiter{\abs}{\lvert}{\rvert}

\begin{document}

\title{Parametric processes in nonlinear structures with reflections: \\
an asymptotic-field approach}

\author{Tadeu Tassis}
\email{tadeu.tassis@polymtl.ca}
\affiliation{D\'epartement de g\'enie physique, \'Ecole polytechnique de Montr\'eal, Montr\'eal, QC, H3T 1J4, Canada}

\author{Salvador Poveda-Hospital}
\affiliation{D\'epartement de g\'enie physique, \'Ecole polytechnique de Montr\'eal, Montr\'eal, QC, H3T 1J4, Canada}
 
\author{Nicol\'as Quesada}
\affiliation{D\'epartement de g\'enie physique, \'Ecole polytechnique de Montr\'eal, Montr\'eal, QC, H3T 1J4, Canada}

\author{Martin Houde}
\affiliation{D\'epartement de g\'enie physique, \'Ecole polytechnique de Montr\'eal, Montr\'eal, QC, H3T 1J4, Canada}

\begin{abstract}
    The generation of engineered quantum states of light via nonlinear processes is fundamental for quantum technologies based on photons.
    Although embedding nonlinear materials within resonant structures allows for the enhancement and tailoring of photon properties, accurately modeling these quantum interactions remains a challenge.
    In this work, we apply the asymptotic-fields formalism, an approach based on scattering theory, to describe nonlinear optical processes within a Fabry-P\'erot cavity.
    Unlike previous applications of this formalism, we explicitly account for reflections in the system.
    We derive the interaction Hamiltonian and calculate photon-pair generation rates using perturbation theory.
    The versatility of this model is illustrated through three examples: (i) spontaneous parametric down-conversion in an idealized cavity with flat-response mirrors; (ii) the generation of counter-propagating photon pairs in a periodically-poled material; and (iii) spontaneous four-wave mixing in a cavity built with Bragg reflectors.
\end{abstract}

\maketitle

\section{Introduction}

The generation of quantum states of light, such as entangled photon pairs, is a cornerstone for optical quantum technologies~\cite{wang2020integrated}.
Quantum nonlinear processes such as spontaneous parametric down-conversion (SPDC) and spontaneous four-wave mixing (SFWM) serve as the main sources of such photon pairs~\cite{quesada2022photon}. 
In these processes, a strong pump generated by a laser interacts with a nonlinear medium and part of the light is converted into light in different frequencies.
In the case of SPDC, we have the conversion of one higher-energy pump photon for two, less energetic ones.
On the other hand in SFWM, two pump photons are exchanged for two new photons, one of higher and one of lower energy than the original pump photons.

The ability to engineer specific properties of the generated photons, such as spectral bandwidth, purity, and generation rates, is essential for applications ranging from optical quantum computing to secure communication and sensing.
A common engineering technique is to place the nonlinear material inside resonant structures, such as Fabry-P\'erot cavities and ring resonators, which allows for resonant enhancement of generation rates and tailoring of spectral properties~\cite{slattery2019background}.
However, accurately modeling nonlinear quantum processes within complex, resonant structures presents a significant challenge~\cite{sorensen2025simple}.

Problems in quantum optics can often be seen as scattering problems: a known input state enters an interaction region and, following the interaction, an output state is measured.
Drawing inspiration from scattering theory in quantum mechanics~\cite{breit1954ingoing}, the authors of Ref.~\cite{liscidini2012asymptotic} propose an asymptotic-field theory to describe nonlinear optical processes.
In this formalism, the fields are expanded in terms of asymptotic-in/out components.
These components are themselves solutions of the linear Maxwell equations, i.e., in the absence of nonlinearity.
In selected cases, the asymptotic fields can be determined using analytical techniques, such as the transfer matrix method.
However, numerical techniques can be employed even in more complex cases.
Once the asymptotic fields are determined, we can easily write the Hamiltonian for the nonlinear interaction in terms of the asymptotic-in/out modes,
enabling the calculation of relevant observables via standard perturbation theory.

The asymptotic-fields formalism has been used to describe nonlinear phenomena in ring resonators in the high-gain regime, in the presence of loss, and with pulsed pump fields~\cite{banic2022two,vendromin2024highly,sloan2025highgain}.
It was also used to study dipole emissions in photonic structures~\cite{macri2025asymptoticfield}.
Although in the latter the authors considered backscattering emissions, reflections by the structures have largely been neglected in studies using this technique.
In this paper, we apply the asymptotic-fields formalism to describe nonlinear processes within a Fabry-P\'erot cavity, explicitly showing how to derive these fields and use them to write the nonlinear Hamiltonian.
As we will see, accounting for reflections is straightforward in this framework.
In order to illustrate our results, we present three applications of the proposed model.
First, we consider the simple case of SPDC in a cavity with flat-response mirrors.
Next, since the formalism naturally incorporates fields propagating in both directions, we investigate the generation of counter-propagating photons in a periodically-poled material.
Finally, we explore the case of SFWM in a cavity built with Bragg reflectors.
This further illustrates the asymptotic-fields formalism as a powerful technique in the theorist's toolbox to model complex photonic systems.

Our choice of system was motivated by a recent work by Sorensen \emph{et al.}~\cite{sorensen2025simple}, where the authors presented a model for SPDC in a cavity based on the transfer-matrix method.
With the asymptotic-fields formalism, we derive a model equivalent to theirs.
Nevertheless, we argue that our approach presents some advantages.
First, it can be readily extended to more complex systems, composed of multiple stacks of nonlinear materials.
Second, it is not restricted to the continuous-wave limit or the undepleted-pump regime.
And, finally, we may be able to describe nonlinear processes that cannot be linearized, such as triplet generation~\cite{lorenaosorio2025strategies}.

The paper is organized as follows.
In Sec.~\ref{sec:general}, we introduce the general asymptotic-in/out formalism, present how to determine the asymptotic fields for a Fabry-P\'erot cavity, and use these to write the interaction Hamiltonian describing SPDC inside the cavity.
Additionally, we use perturbation theory to calculate the photon-pair generation rate in the continuous-wave limit and undepleted-pump regime.
In Sec.~\ref{sec:examples}, we apply the theory we derived to some illustrative examples.
In Sec.~\ref{sec:conclusion}, we present our conclusions.

\section{Asymptotic-field formalism for nonlinear quantum optics} \label{sec:general}

\begin{figure}
    \centering
    \includegraphics[width=\linewidth]{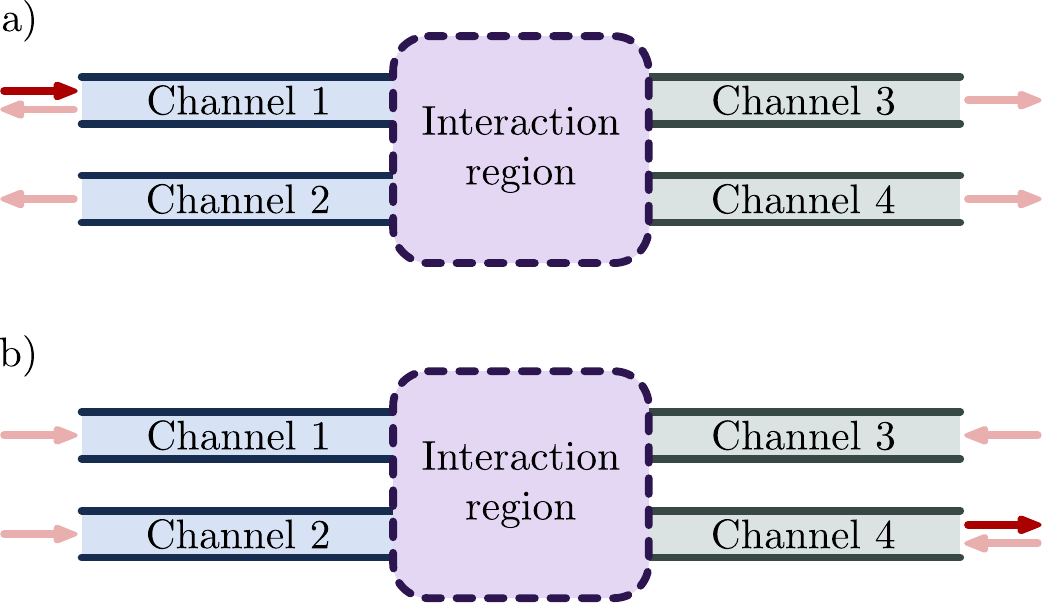}
    \caption{Sketch of a scattering problem in quantum nonlinear optics illustrating the idea of asymptotic-in/out fields. In this example, the system is composed of four channels connected by an interaction region. a) Asymptotic-in field for channel 1. b) Asymptotic-out field for channel 4.}
    \label{fig:asy-fields}
\end{figure}

The idea of asymptotic-in/out states figures prominently in scattering theory~\cite{breit1954ingoing}.
In that context, the asymptotic states are solutions of Schr\"odinger's equation with a particular property.
For very early times, the asymptotic-in state describes a wavepacket moving freely towards the interaction region.
Similarly, asymptotic-out states describe wavepackets moving freely away from the interaction regions for very late times.
It becomes natural then to expand initial states in terms of asymptotic-in components, and final states in terms of asymptotic-out components.

Following this idea, Liscidini, Helt and Sipe~\cite{liscidini2012asymptotic} proposed an asymptotic-fields treatment to nonlinear optics problems.
In this context, the asymptotic fields are solutions of the linear Maxwell's equations, and describe fields moving towards the interaction region for very early times (asymptotic-in), or fields moving away from it for very late times (asymptotic-out).
As an example, consider the system in Fig.~\ref{fig:asy-fields}, consisting of four channels connected to the interaction region.
Fig.~\ref{fig:asy-fields}a presents the asymptotic-in field for channel 1 at any time: we have a field component propagating towards the interaction region in channel 1 and components moving away from it in all the other channels.
For $t \rightarrow -\infty$, the exiting components vanish through destructive interference~\cite{breit1954ingoing}, such that we are left with only the incident part.
Likewise, we present the asymptotic-out field for channel 4 in Fig.~\ref{fig:asy-fields}b.
Here, we have components moving towards the interaction region in all the channels, but we only have a component moving away from it in channel 4.

While in the case of scattering theory the asymptotic states are the solutions for the scattering problem, in the case of nonlinear optics problems, the asymptotic fields are solutions of the linear problem only.
Nevertheless, they form natural bases to expand the fields and write the nonlinear Hamiltonian~\cite{liscidini2012asymptotic}.
Consider the process of SPDC: a pump photon enters the nonlinear material and is converted into two, less energetic photons, called the signal and the idler photons.
In terms of asymptotic fields, then, it becomes natural to write the pump component in terms of asymptotic-in fields, and the converted photons, in terms of asymptotic-out ones.
That is, we expand the displacement field (which is the natural field to use in quantum nonlinear optics~\cite{quesada2017you}) as
\begin{equation}
    \mb{D}(\mb{r}) = \mb{D}_P^\mr{in}(\mb{r}) + \mb{D}_S^\mr{out}(\mb{r}) + \mb{D}_I^\mr{out}(\mb{r}),
    \label{eq:displacement-def}
\end{equation}
where the subscripts, $P, S, I$, label the pump, signal and idler components, respectively.
Each of these components can be, in turn, expanded as
\begin{equation}
    \mb{D}_J^\mr{in(out)} = \sum_X \int dk \, \mb{D}_{JXk}^\mr{in(out)} a_{JXk}^\mr{in(out)} + \hc,
    \label{eq:asy-modes-def}
\end{equation}
where $a_{JXk}^\mr{in(out)}$ annihilates an incoming (outgoing) photon in mode $J=P, S, I$, with vacuum wavenumber $k$, and in channel $X$.

The interaction Hamiltonian for a second-order nonlinearity can be written as~\cite{quesada2022photon}
\begin{equation}
    H_\mr{NL} = - \frac{1}{3 \epsilon_0} \int d^3x \, \Gamma_{2}^{lmn}(\mb{r}) D^l(\mb{r}) D^m(\mb{r}) D^n(\mb{r}),
    \label{eq:hamilt-def}
\end{equation}
where we are summing over repeated indices.
By plugging Eq.~\eqref{eq:asy-modes-def} into \eqref{eq:hamilt-def} and keeping only terms that contribute to SPDC, we find
\begin{align}
    H_\mr{NL} =
    \sum_{X, X'} & \int d^3 k \,
    \Bigl\{ J_{X X'}(k_1, k_2, k_3)
    \nonumber \\
    &\times [a_{S X k_1}^\mr{out}]^\dagger [a_{I X' k_2}^\mr{out}]^\dagger a_{P Y k_3}^\mr{in} \Bigr\} + \hc,
    \label{eq:hamilt-a}
\end{align}
where $d^3k \equiv dk_1 \, dk_2 \, dk_3$ and
\begin{align}
    & J_{X X'}(k_1, k_2, k_3) \equiv
    \nonumber \\
    & \quad - \frac{2}{\epsilon_0} \int d^3x \;
    \Gamma_2^{lmn} \,
    [ D^{\mr{out}, \, l}_{S X k_1} ]^* \,
    [ D^{\mr{out}, \, m}_{I X' k_2} ]^* \,
    D^{\mr{in}, \, n}_{P Y k_3}.
    \label{eq:J-def}
\end{align}
To derive this expression, we have used the fact that $\Gamma_2^{lmn}$ is symmetric in all its indices.
It is also implied that the pump photon enters the interaction region via a single channel, $Y$, and there is no sum over different input channels.

\subsection{Asymptotic modes for a Fabry-P\'erot cavity}
\label{sec:fp}

\begin{figure}[tb]
    \centering
    \includegraphics[width=\linewidth]{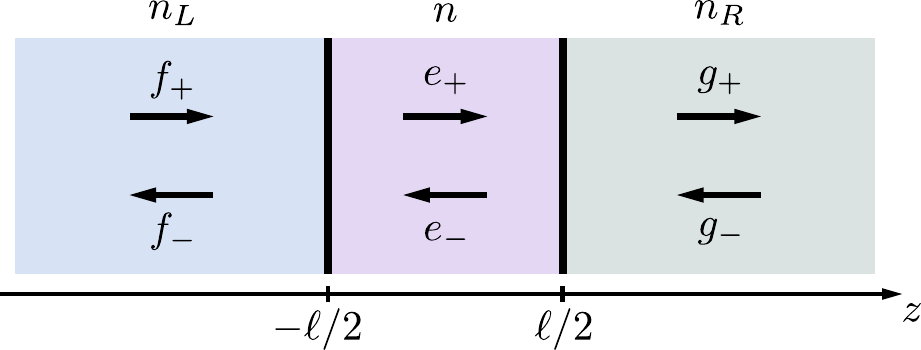}
    \caption{Illustration of a Fabry-P\'erot cavity.
    The left (L) and right (R) channels are connected to a cavity having semi-reflective surfaces at $z = \pm \ell / 2$.
    The field amplitudes $f_{\pm}$ and $e_{\pm}$ are depicted showing the direction of propagation of the associated fields.}
    \label{fig:fp}
\end{figure}

A Fabry-P\'erot cavity [Fig.~\ref{fig:fp}] is constructed by two semi-reflective mirrors.
Like ring resonators, interference effects in the cavity can potentially enhance nonlinear effects when a nonlinear material is placed inside it~\cite{slattery2019background}.
In order to apply the formalism laid out in the last section, we consider the region inside the cavity as the interaction region.
Additionally, we have two channels connected to the cavity, which we label the left (L) and right (R) channels.
In the rest of the text, we consider that the pump source is connected to the left channel.
Thus we can set $Y=L$ in Eqs.~\eqref{eq:hamilt-a} and \eqref{eq:J-def}.

We decompose the asymptotic modes in Eq.~\eqref{eq:asy-modes-def} into transversal and longitudinal terms as
\begin{equation}
    \mb{D}_{J X k}^\mr{in(out)} =
    \sqrt{\frac{\hbar \omega_{J k}}{4 \pi}}
    \mb{d}_{J X k}^\mr{in(out)} (\mb{r}_\perp) \phi_{J X k}^\mr{in(out)}(z).
    \label{eq:asy-modes-ampl}
\end{equation}
Plugging Eq.~\eqref{eq:asy-modes-ampl} into \eqref{eq:J-def}, we find
\begin{equation}
    J_{X X'}(k_1, k_2, k_3) =
    - \sqrt{\frac{\hbar^3 \omega_{S k_1} \omega_{I k_2} \omega_{P k_3}}{16 \pi^3 \epsilon_0^2}} \,
    \Gamma_\perp I_{X X'},
    \label{eq:J-coeff}
\end{equation}
where
\begin{align}
    \Gamma_\perp \equiv &\int d^2x \,
    \Gamma_2^{lmn}(\mb{r}_\perp)
    [d_{S X k_1}^{\mr{out}, l}(\mb{r}_\perp)]^*
    \nonumber \\
    &\times
    [d_{I X' k_2}^{\mr{out}, m}(\mb{r}_\perp)]^*
    d_{P L k_3}^{\mr{in}, n}(\mb{r}_\perp)
\end{align}
and
\begin{equation}
    I_{X X'} \equiv \int_{-\ell/2}^{\ell/2} dz \,
    [\phi_{S X k_1}^\mr{out}(z)]^*
    [\phi_{I X' k_2}^\mr{out}(z)]^*
    \phi_{P L k_3}^\mr{in}(z).
    \label{eq:int-z}
\end{equation}
Notice that we have assumed that the nonlinear tensor, $\Gamma_2^{lmn}$, does not vary considerably in the $z$-direction.
However, if the longitudinal variation can be factorized as $\Gamma_2^{lmn}(\mb{r}) = \tilde{\Gamma}_2^{lmn}(\mb{r_\perp}) \, \gamma(z)$, it is straightforward to include it in the calculation.

We recall that the asymptotic modes, Eq.~\eqref{eq:asy-modes-ampl}, are solutions of the linear theory for the isolated channels.
Suppose that the left and right channels are so long that we can consider the displacement field in the isolated channels to be described by plane waves in the longitudinal direction.
Since reflections play a central role in the problem, we consider the longitudinal term, $f_{JXk}^\mr{in(out)}$, to be a superposition of left- and right-moving waves in each region, i.e.,
\begin{align}
    & \phi_{JXk}^\mr{in(out)} (z) =
    \nonumber \\
    & \quad \begin{cases}
        f_{+, JXk}^\mr{in(out)} e^{i K_L z} + f_{-, JXk}^\mr{in(out)} e^{-i K_L z}, & z < -\ell/2 \\
        e_{+, JXk}^\mr{in(out)} e^{i K z} + e_{-, JXk}^\mr{in(out)} e^{-i K z}, & \abs{z} < \ell/2 \\
        g_{+, JXk}^\mr{in(out)} e^{i K_R z} + g_{-, JXk}^\mr{in(out)} e^{-i K_R z}, & z > \ell/2,
    \end{cases}
    \label{eq:f-def}
\end{align}
where $K_L = n_L(k) k$, $K_R = n_R(k) k$, and $K = n(k) k$ are the material wavenumbers for the left and right channels, and for the nonlinear material inside the cavity, respectively.
Plugging Eq.~\eqref{eq:f-def} into the $z$ integral, Eq.~\eqref{eq:int-z}, we find
\begin{align}
    I_{X X'} & \approx
    \Bigl(
    [e_{+, S X k_1}^\mr{out}]^*
    [e_{+, I X' k_2}^\mr{out}]^*
    e_{+, P L k_3}^\mr{in}
    \Bigr) \int_{-\ell/2}^{\ell/2} dz \, e^{i \Delta K z}
    \nonumber \\
    & \quad + \Bigl(
    [e_{-, S X k_1}^\mr{out}]^*
    [e_{-, I X' k_2}^\mr{out}]^*
    e_{-, P L k_3}^\mr{in}
    \Bigr) \int_{-\ell/2}^{\ell/2} dz \, e^{-i \Delta K z}
    \nonumber \\
    & =  \ell \, \sinc (\tfrac{\Delta K \ell}{2})
    \Bigl(
    [e_{+, S X k_1}^\mr{out}]^*
    [e_{+, I X' k_2}^\mr{out}]^*
    e_{+, P L k_3}^\mr{in}
    \nonumber \\
    & \quad + [e_{-, S X k_1}^\mr{out}]^*
    [e_{-, I X' k_2}^\mr{out}]^*
    e_{-, P L k_3}^\mr{in}
    \Bigr),
    \label{eq:int-z-sinc}
\end{align}
where $\Delta K \equiv K_3 - K_1 - K_2$ and $K_i \equiv K(k_i) = n(k_i) k_i$.

In order to derive Eq.~\eqref{eq:int-z-sinc}, we have neglected terms describing counter-propagating pair generation, i.e., when one or both of the generated photons propagates in the opposite direction to the annihilated pump photon.
In general, these terms present a large phase-mismatch.
Thus, their contribution is much smaller than phase-matched terms.
However, in Sec.~\ref{sec:counter} we consider an example where some of these counter-propagating terms become dominant.

\begin{figure}[tb]
    \centering
    \includegraphics[width=\linewidth]{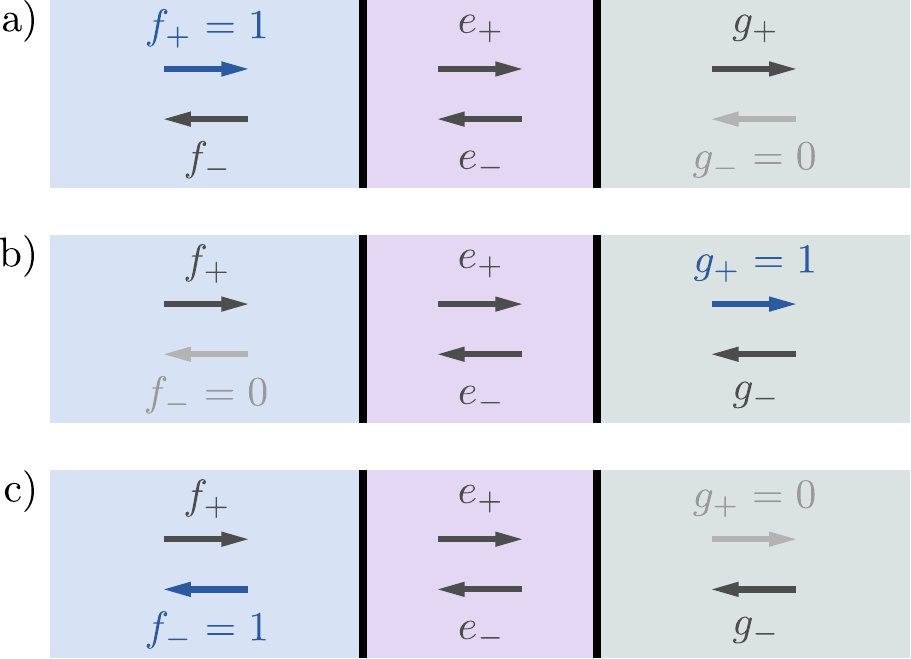}
    \caption{Sketch of the asymptotic conditions on the mode amplitudes.
    a) Asymptotic-in mode from the left; b) asymptotic-out mode from the right; c) asymptotic-out mode from the left.}
    \label{fig:asy-amplitudes}
\end{figure}

In order to relate the amplitudes of Eq.~\eqref{eq:f-def} in the different regions of Fig.~\ref{fig:fp}, we can use the transfer-matrix method~\cite{griffiths2001waves, katsidis2002general}.
Each mirror in the cavity has an associated transfer matrix, $M_i$, $i=1,2$, that transforms the amplitudes in each side of the mirror according to
\begin{equation}
    \begin{pmatrix}
        e_+ \\ e_-
    \end{pmatrix}
    = M_1
    \begin{pmatrix}
        f_+ \\ f_-
    \end{pmatrix}
    \text{ and }
    \begin{pmatrix}
        g_+ \\ g_-
    \end{pmatrix}
    = M_2
    \begin{pmatrix}
        e_+ \\ e_-
    \end{pmatrix}.
    \label{eq:transfer-mat}
\end{equation}
In order to solve the system of equations given in Eq.~\eqref{eq:transfer-mat},
we need to impose the asymptotic conditions of the specific asymptotic mode we are trying to determine.
For the asymptotic-in mode from the left, we expect, at very early times, to have all the fields moving towards the cavity in the left channel and no field approaching from the right.
Thus, we impose the amplitude associated with the right-moving wave in the left channel to be equal to unity, $f_+ = 1$, and the amplitude associated with the left-moving waves in the right channel to be zero, $g_- = 0$.
In the asymptotic-out mode from the right, we expect only right-moving waves in the right channel and no left-moving waves in the left channel for very late times, i.e., $g_+ = 1$ and $f_- = 0$.
Following the same argument, for the asymptotic-out mode from the left we impose $f_- = 1$ and $g_+ = 0$.
These three scenarios are illustrated in Fig.~\ref{fig:asy-amplitudes}.

With the asymptotic conditions described above, we can solve the system in Eq.~\eqref{eq:transfer-mat} and determine the amplitudes inside the cavity, which in turn can be used to calculate the quantity in Eq.~\eqref{eq:int-z-sinc}.
Using a general form for the matrices $M_1$ and $M_2$, we arrive at the following form for the amplitudes $e_+$ and $e_-$,
\begin{align}
    e_{+, JLk}^\mr{in} & = \frac{[(M_1)_{11}(M_1)_{22} - (M_1)_{12}(M_1)_{21}](M_2)_{22}}{(M_1)_{12}(M_2)_{21} + (M_1)_{22}(M_2)_{22}}
    \nonumber, \\
    e_{-, JLk}^\mr{in} & = \frac{[(M_1)_{12}(M_1)_{21} - (M_1)_{11}(M_1)_{22}](M_2)_{21}}{(M_1)_{12}(M_2)_{21} + (M_1)_{22}(M_2)_{22}}
    \nonumber, \\
    e_{+, JRk}^\mr{out} & = \frac{(M_1)_{11}}{(M_1)_{11}(M_2)_{11} + (M_1)_{21}(M_2)_{12}}
    \nonumber, \\
    e_{-, JRk}^\mr{out} & = \frac{(M_1)_{21}}{(M_1)_{11}(M_2)_{11} + (M_1)_{21}(M_2)_{12}}
    \nonumber, \\
    e_{+, JLk}^\mr{out} & = \frac{[(M_1)_{12}(M_1)_{21} - (M_1)_{11}(M_1)_{22}](M_2)_{12}}{(M_1)_{11}(M_2)_{11} + (M_1)_{21}(M_2)_{12}}
    \nonumber, \\
    e_{-, JLk}^\mr{out} & = \frac{[(M_1)_{11}(M_1)_{22} - (M_1)_{12}(M_1)_{21}](M_2)_{11}}{(M_1)_{11}(M_2)_{11} + (M_1)_{21}(M_2)_{12}}.
    \label{eq:asy-ampli}
\end{align}
In Sec.~\ref{sec:examples} we will analyze specific systems and will provide examples of transfer matrices in each of them.

\subsection{Photon-pairs in the undepleted-pump regime and continuous-wave limit}
\label{sec:cw-regime}

In order to calculate the photon-pair generation rates in the Fabry-P\'erot cavity, we make two simplifying assumptions.
The first is that the pump field is so strong that it is not disturbed by the interaction and can be treated classically, this is called the undepleted-pump regime.
The second is the continuous-wave limit, where we consider the spatial distribution of the pump field to be much larger that any other length scale in the problem, such that it has a very sharp, well-defined wavenumber~\cite{ou2017}.
Considering these assumptions, we can make the map
\begin{equation}
    a_{P L k_3}^\mr{in} \mapsto \alpha_P \, \delta(k_3 - k_P),
    \label{eq:undep-pump}
\end{equation}
where $\alpha_P$ is a c-number and $k_P$ is the pump vacuum wavenumber.
We substitute Eq.~\eqref{eq:undep-pump} into the Hamiltonian, Eq.~\eqref{eq:hamilt-a}, writing it first in the interaction picture,
\begin{align}
    \tilde{H}_\mr{NL}(t) =
    &\sum_{X, X'} \int d^3 k \,
    \Bigl\{ J_{X X'}(k_1, k_2, k_3)
    \nonumber \\
    &\times [a_{S X k_1}^\mr{out}]^\dagger [a_{I X' k_2}^\mr{out}]^\dagger a_{P L k_3}^\mr{in} e^{-i \Omega t} \Bigr\} + \hc
    \\
    \mapsto &\sum_{X, X'} \int d^2 k \,
    \Bigl\{ J_{X X'}(k_1, k_2, k_P)
    \nonumber \\
    &\times [a_{S X k_1}^\mr{out}]^\dagger [a_{I X' k_2}^\mr{out}]^\dagger \alpha_P \, e^{-i \Omega t} \Bigr\} + \hc,
    \label{eq:hamilt-undep}
\end{align}
where $\Omega \equiv \omega_{P k_P} - \omega_{S k_1} - \omega_{I k_2}$.

In order to calculate the state of the system after the interaction, we assume that the nonlinearity is weak.
This implies that the probability of conversion of pump photons is small.
Moreover, we consider that the signal and idler fields are initially in the vacuum state.
According to first-order perturbation theory, if the nonlinear interaction takes place during an interval $T$, the state of the system after the interaction will be
\begin{equation}
    \ket{\Psi} \approx \ket{\vac} + \veps \ket{\mr{II}}, 
\end{equation}
where $\abs{\veps}^2 \ll 1$ and $\ket{\mr{II}}$ is a two-photon state such that
\begin{equation}
    \veps \ket{\mr{II}} = - \frac{i}{\hbar} \int_{-T/2}^{T/2} dt \, \tilde{H}_\mr{NL}(t) \ket{\vac}.
\end{equation}
Substituting Eq.~\eqref{eq:hamilt-undep} into this last equation gives us
\begin{align}
    \veps \ket{\mr{II}} & =
    - \frac{i}{\hbar} \sum_{X, X'} \int d^2k
    \left( \int_{-T/2}^{T/2} dt \, e^{-i \Omega t} \right)
    \nonumber \\
    & \quad \times J_{X X'} \, \alpha_P \, [a_{S X k_1}^\mr{out}]^\dagger [a_{I X' k_2}^\mr{out}]^\dagger \ket{\vac}
    \nonumber \\
    & = - \frac{i}{\hbar} \sum_{X, X'} \int d^2k \,
    T \sinc(\tfrac{\Omega T}{2})
    \nonumber \\
    & \quad \times J_{X X'} \, \alpha_P \, [a_{S X k_1}^\mr{out}]^\dagger [a_{I X' k_2}^\mr{out}]^\dagger \ket{\vac}.
    \label{eq:two-photon-state}
\end{align}

With the states of the generated photons, Eq.~\eqref{eq:two-photon-state}, at hand, we are able to calculate some quantities of interest.
The probability of generating a signal photon with wavenumber $k_1$ that leaves the cavity at channel $X$ and an idler photon with wavenumber $k_2$ that comes out channel $X'$ is
\begin{align}
    P_{SI}(X k_1; X' k_2)
    & \equiv \abs{\braket{S X k_1; I X' k_2 | \Psi}}^2
    \nonumber \\
    & = \frac{T^2 \abs{\alpha_P}^2}{\hbar^2} \, \sinc^2(\tfrac{\Omega T}{2}) \abs{J_{X X'}}^2,
\end{align}
where $\ket{S X k_1; I X' k_2} \equiv [a_{S X k_1}^\mr{out}]^\dagger [a_{I X' k_2}^\mr{out}]^\dagger \ket{\vac}$.
Considering the interaction time is much larger than the inverse frequency difference, $T \Omega \gg 1$, we can take the limit
\begin{equation}
    T \sinc^2(\tfrac{\Omega T}{2}) \rightarrow 2 \pi \delta(\Omega),
\end{equation}
such that
\begin{equation}
    P_{SI}(X k_1; X' k_2) = \frac{2 \pi  \abs{\alpha_P}^2 T}{\hbar^2} \abs{J_{X X'}}^2 \delta(\Omega).
\end{equation}
We can divide this probability by the remaining $T$ factor to arrive at the joint photon-pair generation rate,
\begin{equation}
    R_{SI}(X k_1; X' k_2) = \frac{2 \pi \abs{\alpha_P}^2}{\hbar^2} \abs{J_{X X'}}^2 \delta(\Omega).
    \label{eq:gen-rate}
\end{equation}

With Eq.~\eqref{eq:gen-rate} at hand, we can calculate the spectral distribution of the signal photon by integrating over the idler wavenumbers
\begin{equation}
    S_{X X'}(k_1) = \int dk_2 \frac{2 \pi \abs{\alpha_P}^2}{\hbar^2} \abs{J_{X X'}}^2 \delta(\Omega).
\end{equation}
In terms of wavenumbers, we can write $\Omega = c(k_P - k_1 - k_2)$.
Thus, using $\delta(ax) = \delta(x) / \abs{a}$,
\begin{equation}
    S_{X X'}(k_1) = \frac{2 \pi \abs{\alpha_P}^2}{c \hbar^2} \abs{J_{X X'}(k_1, k_P-k_1, k_P)}^2.
    \label{eq:spectral-rate}
\end{equation}
By integrating over the signal wavenumbers, we arrive at the total photon-pair generation rate
\begin{equation}
    R_{X X'} = \frac{2 \pi \abs{\alpha_P}^2}{c \hbar^2} \int dk_1 \, \abs{J_{X X'}(k_1, k_P-k_1, k_P)}^2.
    \label{eq:total-rate}
\end{equation}

\section{Examples} \label{sec:examples}

In this section we will explore some examples that can be described by the model derived in the previous sections.

\subsection{Fabry-P\'erot cavity with flat mirrors}
\label{sec:flat}

As a first example, we consider that the cavity is built from idealized mirrors having a flat frequency-response.
The transfer matrix for a flat-response mirror placed at $z=0$ is given by~\cite{katsidis2002general}
\begin{equation}
    \tilde{M}^\mr{flat} = \frac{1}{t}
    \begin{pmatrix}
        1 & r \\
        r & 1
    \end{pmatrix},
    \label{eq:tm-flat-0}
\end{equation}
where the reflection and transmission amplitudes, $r$ and $t$, are related by $\abs{r}^2 + \abs{t}^2 = 1$.
If the mirror is placed at $z=a$, we need to transform the amplitudes in Eq.~\eqref{eq:transfer-mat} as
\begin{align}
    \begin{pmatrix}
        \tilde{e}_+ \\ \tilde{e}_-
    \end{pmatrix}
    & = U(a)
    \begin{pmatrix}
        e_+ \\ e_-
    \end{pmatrix}
    \nonumber \\
    & =
    \begin{pmatrix}
        e^{i K a} & 0 \\
        0 & e^{-i K a}
    \end{pmatrix}
    \begin{pmatrix}
        e_+ \\ e_-
    \end{pmatrix},
\end{align}
where $\tilde{e}_i$ are the amplitudes in a shifted frame and $K$ is the material wavenumber.
In this shifted frame, the mirror is located at $\tilde{z} = 0$, such that we can use Eq.~\eqref{eq:tm-flat-0}.
To find the transfer matrix in the displaced reference frame, i.e., the matrix that acts on the amplitudes $e_i$, we use the transformation $U(a)$,
\begin{align}
    M^\mr{flat}(a) & = U^\dagger(a) \tilde{M}^\mr{flat} U(a)
    \nonumber \\
    & = \frac{1}{t}
    \begin{pmatrix}
        1 & r e^{-2i K a} \\
        r e^{2i K a} & 1
    \end{pmatrix}.
\end{align}
Thus, if the cavity illustrated in Fig.~\ref{fig:fp} is composed of flat-response mirrors, we can use $M_1 = M^\mr{flat}(-\ell/2)$ and $M_2 = M^\mr{flat}(\ell/2)$ in Eq.~\eqref{eq:asy-ampli} to determine the particular asymptotic amplitudes,
\begin{align}
  e_{+, JLk}^\mr{in} & = \frac{t_1}{r_1 r_2 e^{2 i k \ell} + 1}
    \nonumber, \\
    e_{-, JLk}^\mr{in} & = - \frac{t_1 r_2 e^{i k \ell}}{r_1 r_2 e^{2 i k \ell} + 1}
    \nonumber, \\
    e_{+, JRk}^\mr{out} & = \frac{t_2 e^{2 i k \ell}}{r_1 r_2 + e^{2 i k \ell}}
    \nonumber, \\
    e_{-, JRk}^\mr{out} & = \frac{r_1 t_2 e^{i k \ell}}{r_1 r_2 + e^{2 i k \ell}}
    \nonumber, \\
    e_{+, JLk}^\mr{out} & = - \frac{t_1 r_2 e^{i k \ell}}{r_1 r_2 + e^{2 i k \ell}}
    \nonumber, \\
    e_{-, JLk}^\mr{out} & = \frac{t_1 e^{2 i k \ell}}{r_1 r_2 + e^{2 i k \ell}}.
    \label{eq:asy-ampli-flat}
\end{align}
These coefficients are equivalent to the ones given in Ref.~\cite{sorensen2025simple}, up to sign and phase conventions.

\begin{figure}[tb]
    \centering
    \includegraphics[width=\linewidth]{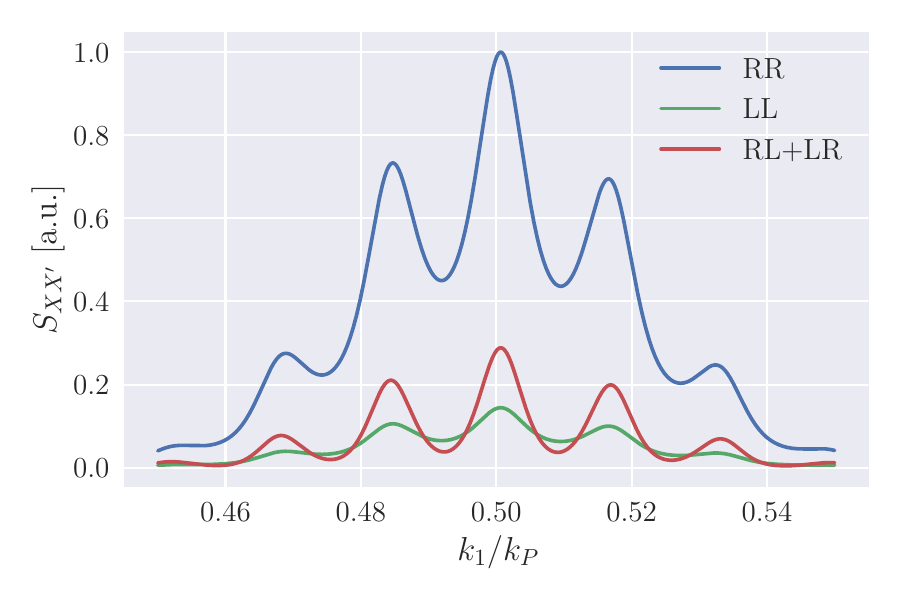}
    \caption{Spectral distribution of signal photons as a function of the signal wavenumber for different output-channel configurations.
    The results were normalized by the maximum rate value.
    The parameters used were: $\lambda_P = 750\mr{nm}$, $r_1 = -r_2 = 0.3$, $\ell = 10.15 \mu \mr{m}$, $(\bar{n}_P, \bar{n}_S, \bar{n}_I) = (2.18, 2.14, 2.22)$, $(\bar{n}_{G,P}, \bar{n}_{G,S}, \bar{n}_{G,I}) = (2.28, 2.18, 2.27)$.}
    \label{fig:spdc-spectral}
\end{figure}

In Fig.~\ref{fig:spdc-spectral}, we present the spectrum of generated photons for a cavity of length $\ell = 10.15 \mu \mr{m}$, reflection amplitudes $r_1 = -r_2 = 0.3$, and for a pump wavelength of $\lambda_P = 750 \mr{nm}$, which was calculated using Eq.~\eqref{eq:spectral-rate}.
The rates were calculated for the different possibilities of output channels: in RR (blue line) both photons are detected in the right channel; in LL (green line) both are detected in the left channel; and in RL+LR (red line) we have each photon leaving the cavity via a different channel.
In order to compare the results with the ones presented in Ref.~\cite{sorensen2025simple}, we multiplied the rate in Eq.~\eqref{eq:spectral-rate} by an envelope function $\eta(k_1)$ to model detectors with nonflat frequency-response, as they had in their experiment.
We considered a Gaussian envelope of the form
\begin{equation}
    \eta(k_1) = e^{- (k_1 - k_0)^2 / (2 \sigma^2)},
\end{equation}
where $k_0 = k_P/2$ and $\sigma = 0.04 k_0$.

Since $\eta$ restricts the bandwidth of generated photons, we can consider the material dispersion to be approximately linear,
\begin{equation}
    K(k) \approx \bar{n} k + \bar{n}_G (k - \bar{k}),
\end{equation}
where $\bar{k}$ is a reference wavenumber, $\bar{n} \equiv n(\bar{k})$ is the refractive index at $\bar{k}$, and $\bar{n}_G \equiv n_G(\bar{k})$ is the group index at $\bar{k}$.
For SPDC, we consider $\bar{k}_P = k_P$ and $\bar{k}_S = \bar{k}_I = k_P / 2$.
In this specific example, we used $(\bar{n}_P, \bar{n}_S, \bar{n}_I) = (2.18, 2.14, 2.22)$ and $(\bar{n}_{G,P}, \bar{n}_{G,S}, \bar{n}_{G,I}) = (2.28, 2.18, 2.27)$, which correspond roughly to index values for bulk $\mr{LiNbO_3}$ at $\lambda_P = 750\mr{nm}$.
We also considered type-II phase matching, i.e., we used indices for the extraordinary polarization for the pump and signal photons, and ordinary polarization for the idler.

\begin{figure}[tb]
    \centering
    \includegraphics[width=\linewidth]{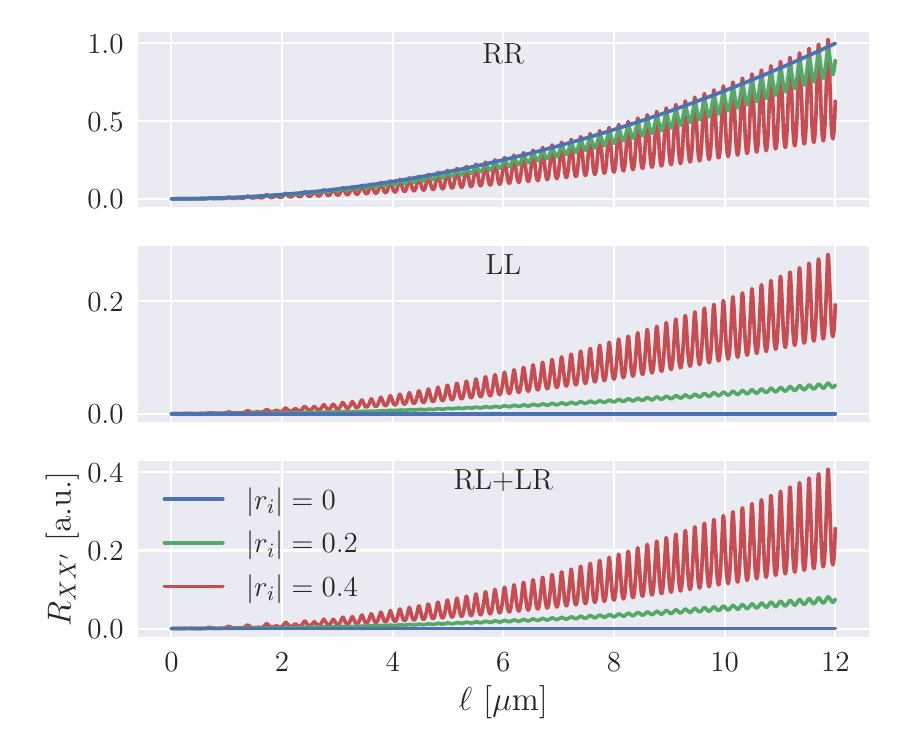}
    \caption{Total rate of pair generation as a function of the cavity length for different reflection amplitudes.
    The results were normalized by the maximum rate value.
    Each panel presents the results for different output-channel configuration.
    The parameters used were: $\lambda_P = 750\mr{nm}$, $(\bar{n}_P, \bar{n}_S, \bar{n}_I) = (2.18, 2.14, 2.22)$, $(\bar{n}_{G,P}, \bar{n}_{G,S}, \bar{n}_{G,I}) = (2.28, 2.18, 2.27)$.}
    \label{fig:spdc-total}
\end{figure}

In Fig.~\ref{fig:spdc-total} we present the (normalized) total rate of photon-pair generation, given by Eq.~\eqref{eq:total-rate}, as a function of the cavity length.
Additionally, we compare rates calculated for different reflection amplitudes: $r_1 = -r_2 = 0$ (blue line), $r_1 = -r_2 = 0.2$ (green line), and $r_1 = -r_2 = 0.4$ (red line).
Notice that even though the resonance effects do not seem to enhance significantly the generation of photons in the RR-output, the rate in the other output configurations increases considerably.
It is also worth mentioning that all other parameters, besides the reflection amplitudes, were kept equal in all three cases.
Thus, in the cases with higher reflection, a larger percentage of the pump is reflected in the first mirror and never enters the cavity.
If the pump power is adjusted accordingly, we expect even further enhancement in the generation rates.
The values for the other parameters were the same as the previous example, Fig.~\ref{fig:spdc-spectral}.

\subsection{Counter-propagating generation in a periodically-poled material}
\label{sec:counter}

Next, we will explore another possible application of the model we presented in Sec.~\ref{sec:general}.
In order to include reflection effects in our model, we had to consider fields propagating in both directions.
However, the inclusion of these terms is also essential if we want to consider the possibility of the material to generate counter-propagating photons.
That is, the scenario where one or both of the generated photons propagate in opposite direction of the input pump photon.
Generally this process is heavily suppressed except for very thin layers of nonlinear materials, since the phase-mismatch is large.
An interesting possibility to enhance it, is the use of a periodically-poled material to achieve quasi-phase-matching (QPM) in these counter-propagating processes~\cite{booth2002counterpropagating,luo2020counterpropagating,chiriano2024purifying}.

In order to include the periodic poling in our model, we return to Eq.~\eqref{eq:J-coeff}, where we factorized the spatial integral.
We consider now that the nonlinear tensor can be written as $\Gamma_2^{lmn}(\mb{r}) = \tilde{\Gamma}_2^{lmn}(\mr{r}_\perp) \gamma(z)$, such that the $z$-integral becomes
\begin{equation}
    I_{X X'} = \int dz \,
    \gamma(z)
    [\phi_{S X k_1}^\mr{out}(z)]^*
    [\phi_{I X' k_2}^\mr{out}(z)]^*
    \phi_{P L k_3}^\mr{in}(z).
    \label{eq:int-z-ppln}
\end{equation}
To achieve QPM, we periodically alternate the orientation of the crystal axis, which has the effect of flipping the sign of the nonlinear coefficient.
A simple model for that is to consider that the coefficient $\gamma$ has a square wave form~\cite{boyd2020nonlinear},
\begin{equation}
    \gamma(z) = \mr{sign}\biggl[ \sin \biggl(\frac{2 \pi (z - z_0)}{\Lambda_\mr{PP}}\biggr) \biggr],
\end{equation}
which alternates sign every half-period, $\Lambda_\mr{PP} / 2$.
It is possible to show that, to achieve QPM, we set the polling period to~\cite{boyd2020nonlinear}
\begin{equation}
  \Lambda_\mr{PP} = \frac{2 \pi}{\Delta \bar{K}},
\end{equation}
where $\Delta \bar{K}$ is the phase-mismatch evaluated at the central frequencies.

When deriving Eq.~\eqref{eq:int-z-sinc} we have neglected the terms that describe counter-propagating generation, as they did not contribute significantly to the problems considered before.
However, they are crucial for the scenario we are considering now.
Thus, plugging Eq.~\eqref{eq:f-def} into Eq.~\eqref{eq:int-z-ppln} and keeping all the terms, we find
\begin{align}
    & I_{X X'} =
    \nonumber \\
    & \quad \Bigl(
    [e_{+, S X k_1}^\mr{out}]^*
    [e_{+, I X' k_2}^\mr{out}]^*
    e_{+, P L k_3}^\mr{in}
    \Bigr) \int dz \, \gamma(z) e^{i \Delta K z}
    \nonumber \\
    & \quad + \Bigl(
    [e_{-, S X k_1}^\mr{out}]^*
    [e_{-, I X' k_2}^\mr{out}]^*
    e_{-, P L k_3}^\mr{in}
    \Bigr) \int dz \, \gamma(z) e^{-i \Delta K z}
    \nonumber \\
    & \quad + \Bigl(
    [e_{-, S X k_1}^\mr{out}]^*
    [e_{+, I X' k_2}^\mr{out}]^*
    e_{+, P L k_3}^\mr{in}
    \Bigr) \int dz \, \gamma(z) e^{i \Delta K_1 z}
    \nonumber \\
    & \quad + \Bigl(
    [e_{+, S X k_1}^\mr{out}]^*
    [e_{-, I X' k_2}^\mr{out}]^*
    e_{-, P L k_3}^\mr{in}
    \Bigr)  \int dz \, \gamma(z) e^{-i \Delta K_1 z}
    \nonumber \\
    & \quad + \Bigl(
    [e_{+, S X k_1}^\mr{out}]^*
    [e_{-, I X' k_2}^\mr{out}]^*
    e_{+, P L k_3}^\mr{in}
    \Bigr) \int dz \, \gamma(z) e^{i \Delta K_2 z}
    \nonumber \\
    & \quad + \Bigl(
    [e_{-, S X k_1}^\mr{out}]^*
    [e_{+, I X' k_2}^\mr{out}]^*
    e_{-, P L k_3}^\mr{in}
    \Bigr) \int dz \, \gamma(z) e^{-i \Delta K_2 z}
    \nonumber \\
    & \quad + \Bigl(
    [e_{-, S X k_1}^\mr{out}]^*
    [e_{-, I X' k_2}^\mr{out}]^*
    e_{+, P L k_3}^\mr{in}
    \Bigr) \int dz \, \gamma(z) e^{i \Delta K_{12} z}
    \nonumber \\
    & \quad + \Bigl(
    [e_{+, S X k_1}^\mr{out}]^*
    [e_{+, I X' k_2}^\mr{out}]^*
    e_{-, P L k_3}^\mr{in}
    \Bigr) \int dz \, \gamma(z) e^{-i \Delta K_{12} z},
    \label{eq:int-z-counter}
\end{align}
where the phase-mismatch terms are as follows: $\Delta K \equiv K_3 - K_1 - K_2$, $\Delta K_1 \equiv K_3 + K_1 - K_2$, $\Delta K_2 \equiv K_3 - K_1 + K_2$, and $\Delta K_{12} \equiv K_3 + K_1 + K_2$.
Again, the material wavenumbers are defined as $K_i \equiv K(k_i) = n(k_i) k_i$.

\begin{figure}[tb]
    \centering
    \includegraphics[width=\linewidth]{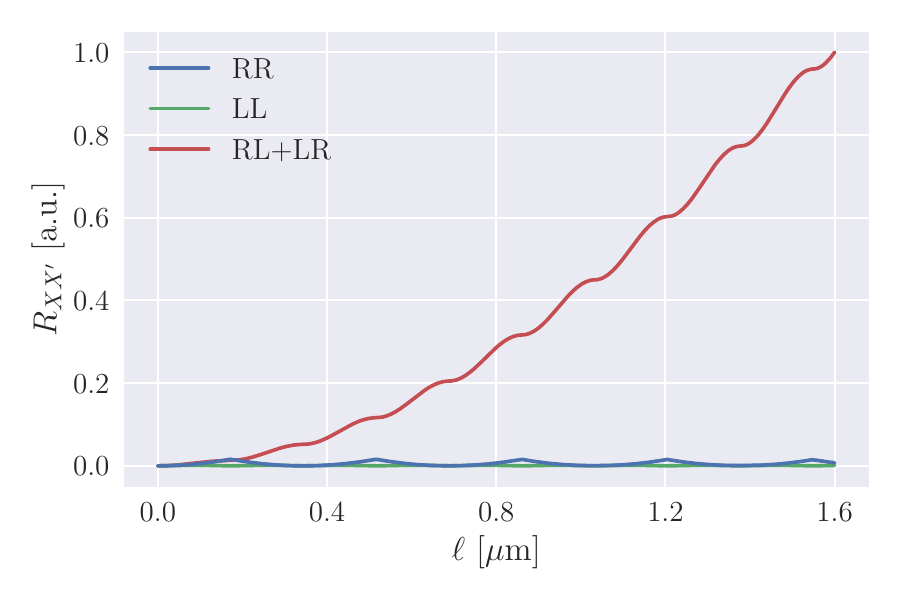}
    \caption{Total rate of pair generation for a periodically-poled material as a function of the material length.
    The poling period was set to achieve QPM when the generated photons propagate in opposite directions to each other.
    The results were normalized by the maximum rate value.
    The parameters used were: $\lambda_P=750\mr{nm}$, $(\bar{n}_P, \bar{n}_S, \bar{n}_I) = (2.18, 2.14, 2.14)$ and $(\bar{n}_{G,P}, \bar{n}_{G,S}, \bar{n}_{G,I}) = (2.28, 2.18, 2.18)$.}
    \label{fig:ppln-counter-total}
\end{figure}

To calculate the rate of photon-pair generation in a periodically-poled material and considering the possibility of generating counter-propagating photons, we use Eq.~\eqref{eq:int-z-ppln} to calculate Eq.~\eqref{eq:total-rate}.
In Fig.~\ref{fig:ppln-counter-total} we present the total rate of pair generation for a crystal prepared to achieve QPM for counter-propagating signal-idler pairs.
Specifically, we set the poling-period to
\begin{equation}
  \Lambda_\mr{PP} = \frac{2 \pi}{\Delta \bar{K}_2}.
\end{equation}
Note that, since we are assuming for simplicity that the signal and idler modes are equal and that their central frequencies are half the pump frequency, we have $\Delta \bar{K}_2 = \bar{K}_1$.
In Eq.~\eqref{eq:int-z-counter}, then, terms with both phase-mismatches will contribute.
We considered a pump at $\lambda_P = 750 \mr{nm}$, and for the reference refractive and group indices we used, respectively, $(\bar{n}_P, \bar{n}_S, \bar{n}_I) = (2.18, 2.14, 2.14)$ and $(\bar{n}_{G,P}, \bar{n}_{G,S}, \bar{n}_{G,I}) = (2.28, 2.18, 2.18)$.

\subsection{SFWM in a cavity with Bragg reflectors}
\label{sec:bragg}

Finally, we consider another nonlinear process used to generate photon pairs: SFWM.
This is a third-order process in which two pump photons are annihilated to generate signal-idler pairs such that $2 \omega_P = \omega_S + \omega_I$.
We also consider the more complex case of a cavity formed by Bragg reflectors~\cite{xie2020onchip,kellner2025low}.
These are built with a grating technique, where several thin layers of material with slightly different refractive indices are alternately stacked.
The combined interference over the multiple layers has the net effect of only a small bandwidth being significantly reflected by the structure, while the majority of the spectrum is almost completely transmitted.
Analogously to what was done for SPDC in Sec.~\ref{sec:general}, we can apply the asymptotic-fields formalism to SFWM.
The results are summarized in Appendix~\ref{sec:sfwm}.

\begin{figure}[tb]
    \centering
    \includegraphics[width=\linewidth]{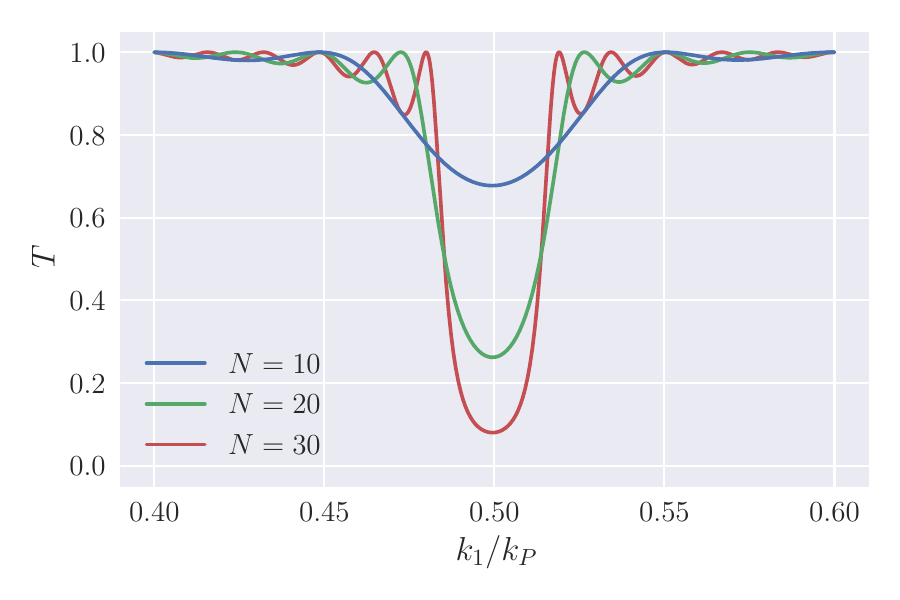}
    \caption{Transmission coefficients of Bragg reflectors built with different number of layers as a function of the wavenumber.
    The parameters used were: $\lambda_P = 750\mr{nm}$, $n_1 = 1.5$, $n_2=1.6$, and $\Lambda_\mr{BG} = 2 \pi / (n_\mr{eff} k_P)$.}
    \label{fig:bg-transmission}
\end{figure}

As mentioned previously, the Bragg grating is formed by a stack of alternating layers with close refractive indices, $n_1$ and $n_2$~\cite{griffiths2001waves, yeh2005optical}.
We can derive the total transfer matrix for a Bragg reflector as follows.
Consider that the interface between two media, $1$ and $2$, is at $z=a$.
Using the boundary conditions imposed by Maxwell's equations on the displacement field, it is possible to show that the transfer matrix from medium $1$ to $2$ is given by~\cite{griffiths2001waves}
\begin{align}
    & M^\mr{12}(k, a) =
    \nonumber \\
    & \; \frac{1}{2 \rho^2}
    \begin{pmatrix}
        (1+\rho) e^{i(K_1 - K_2)a} & (1-\rho) e^{-i(K_1 + K_2)a} \\
        (1-\rho) e^{i(K_1 - K_2)a} & (1+\rho) e^{-i(K_1 + K_2)a}
    \end{pmatrix},
\end{align}
where $K_i = n_i k$, $i=1,2$, is the material wavenumber for the left (right) side of the interface, and $\rho \equiv n_1 / n_2$.
When going from medium $2$ to $1$, we can write an analogous matrix $M^{21}$.
So, the total transfer matrix for a Bragg reflector of grating period $\Lambda_\mr{BG}$, $N$ layers and starting at $z=z_0$ is
\begin{align}
    & M^\mr{BG}(k, \Lambda_\mr{BG}, N, z_0) =
    \nonumber \\
    & \quad \prod_{m=0}^{N}
    M^{21}(k, z_m + \Lambda_\mr{BG} / 2)
    M^{12}(k, z_m),
\end{align}
where $z_m \equiv z_0 + 2m \Lambda_\mr{BG} / 2$ and the product is ordered such that terms with larger $z_m$ are multiplied to the left of previous terms.

For an arbitrary transfer matrix $M$, the transmission coefficient, i.e., the percentage of the intensity transmitted by the structure is given by~\cite{griffiths2001waves}
\begin{equation}
    T(k) = \abs*{\frac{M_{11}(k) M_{22}(k)-M_{12}(k) M_{21}(k)}{M_{11}(k)}}^2.
\end{equation}
It is possible to show that the transmission coefficient is minimum (maximum reflection) for~\cite{yeh2005optical}
\begin{equation}
    k_0 = \frac{\pi}{n_\mr{eff} \Lambda_\mr{BG}},
\end{equation}
where $n_\mr{eff} \equiv 2 n_1 n_2 / (n_1 + n_2)$.
In Fig.~\ref{fig:bg-transmission} we present the transmission coefficient for a Bragg reflector with refractive indices $n_1 = 1.5$ and $n_2 = 1.6$, and grating period $\Lambda_\mr{BG} = 2 \pi / (n_\mr{eff} k_P)$ for a pump at $\lambda_P = 750nm$.
We compare the transmission for different number of layers.
The more layers, the more accentuated is the dip in transmission for the center wavenumber, $k_P / 2$ in this example.

\begin{figure}[tb]
    \centering
    \includegraphics[width=\linewidth]{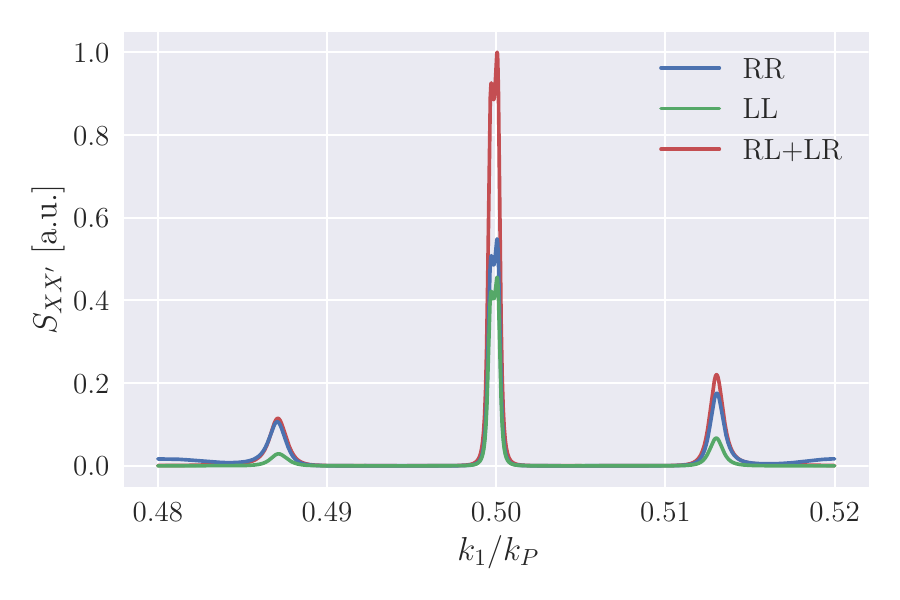}
    \caption{Spectral distribution of signal photons generated via SFWM in a Bragg cavity.
    The results were normalized by the maximum rate value.
    The parameters used were: $\lambda_P = 750 \mr{nm}$, $\ell = 9.98 \mu\mr{m}$, $n_1 = 1.5$, $n_2 = 1.6$, $\Lambda_\mr{BG} = \lambda_P / n_\mr{eff}$, $N=30$, $(\bar{n}_P, \bar{n}_S, \bar{n}_I) = (2.18, 2.14, 2.22)$, and $(\bar{n}_{G,P}, \bar{n}_{G,S}, \bar{n}_{G,I}) = (2.28, 2.18, 2.27)$.}
    \label{fig:bg-spectral}
\end{figure}

A Fabry-P\'erot cavity built with Bragg reflectors allows tuning of the reflected frequencies.
This in turn allows for the generation of very narrowband signal photons~\cite{abolghasem2009bandwidth,fekete2013ultranarrowband,monteiro2014narrowband}.
For that, the mirrors are tuned to a certain frequency and the length is chosen to increase resonance effects inside the cavity.
To illustrate this application, we present the spectral generation rates for a cavity built with Bragg reflectors in Fig.~\ref{fig:bg-spectral}.
The generation rate calculated for SFWM using the asymptotic-fields formalism is given in Eq.~\eqref{eq:rate-sfwm}.
We considered mirrors with $N=30$ layers and the grating period was chosen to have maximum reflection at half the pump wavenumber, $k_P/2$.
The cavity length was set to $\ell = 9.98 \mu\mr{m}$.
As in Sec.~\ref{sec:flat}, the dispersion relation in the material was considered to be approximately linear with reference refractive and group indices set, respectively, to $(\bar{n}_P, \bar{n}_S, \bar{n}_I) = (2.18, 2.15, 2.19)$ and $(\bar{n}_{G,P}, \bar{n}_{G,S}, \bar{n}_{G,I}) = (2.28, 2.18, 2.32)$, such that we have phase-matching for $K_S = K_P/2$ and $K_I = 3 K_P/2$.

\section{Final remarks}
\label{sec:conclusion}

In this work we used the asymptotic-fields formalism to study nonlinear processes in a Fabry-P\'erot cavity.
Expanding the fields in terms of their asymptotic-in/out components, we derived the nonlinear Hamiltonian that describes SPDC and SFWM inside the cavity.
Using these results, and considering the pump field to be undepleted by the interaction and in the continuous-wave regime, we derived expressions for the spectral and total rate of photon-pair generation.

To demonstrate the flexibility of the model we presented, we applied the results to some illustrative examples.
First, we considered SPDC in an idealized cavity with flat-response mirrors.
The results we obtained correctly capture the effects of resonant enhancement of photon-pair generation in a cavity.
Next, we showed that the theory naturally accommodates the process of generation of counter-propagating photons.
In order to enhance this process, we used a periodically poled material to achieve quasi-phase-matching for photon-pairs propagating in opposite directions.
Finally, we considered a different nonlinear process for generation of photon pairs: SFWM.
In this case, we showed the potential of the asymptotic-fields method to describe more complex components by applying it to a cavity built with Bragg reflectors.
As an example of application of such system, we presented the generation narrowband photon-pairs.

Looking ahead, we believe there are several promising directions for future
study. Although we focused on the undepleted-pump regime, the formalism is not
restricted to this limit and it is possible to extend it to high-gain
regimes~\cite{vendromin2024highly}. Similarly, even though we neglected losses
in our system, loss effects can be incorporated naturally into the
asymptotic-fields formalism~\cite{banic2022two}. Finally, one direction that we
believe to be very promising is the study of nonlinear processes in the case of
a broadband pump field, breaking the CW limit~\cite{ou2007}. Although this has been studied
in the context of microring resonators~\cite{sloan2025highgain}, exploring how
reflection effects influence the temporal and spectral properties of a finite
pulse could lead to interesting new insights. A simple way to first approach
this problem in the undepleted-pump regime is to substitute the transformation
in Eq.~\eqref{eq:undep-pump}. Instead of using a sharp Dirac delta, one could
consider a smooth function, such as a Gaussian. This would lead to an
additional integral over the pump wavenumbers in the rates calculated in
Sec.~\ref{sec:cw-regime} and, possibly, richer spectral properties for the
generated photons.

The asymptotic-fields approach provides a powerful tool in the study of complex photonic structures.
While we focused on the case of photon-pair generation in a Fabry-P\'erot cavity, there is a clear methodology to deal with more intricate structures and other nonlinear processes.
The ability to derive accurate theoretical models and to simplify the analysis in complex systems has great potential for the development and enhancement of optical devices for, e.g., quantum information and sensing applications.

\section*{Acknowledgements}
The authors thank C. Vendromin and J.E. Sipe for insightful discussions. They acknowledge financial support from the Natural Science and Engineering Research Council of Canada  Quantum Consortium Program (Quantamole).

\section*{Data availability}
The numerical work presented here can be reproduced using the code in our GitHub repository (\url{https://github.com/polyquantique/fabry_perot_asy}).

\appendix

\section{Asymptotic-fields for SFWM in a cavity} \label{sec:sfwm}

In Sec.~\ref{sec:general}, we calculated the nonlinear Hamiltonian that describes SPDC in a Fabry-P\'erot cavity.
Here, we follow the same recipe to derive the Hamiltonian describing SFWM in a cavity.

Instead of Eq.~\eqref{eq:hamilt-def}, we start from the Hamiltonian for a third-order nonlinearity,
\begin{align}
    & H_\mr{NL}^\mr{SFWM} =
    \nonumber \\
    & \quad - \frac{1}{4 \epsilon_0} \int d^3x \, \Gamma_{3}^{lmno}(\mb{r}) D^l(\mb{r}) D^m(\mb{r}) D^n(\mb{r}) D^o(\mb{r}).
    \label{eq:hamilt-3rd}
\end{align}
By plugging Eqs.~\eqref{eq:displacement-def} and \eqref{eq:asy-modes-def} and keeping only terms that contribute to SFWM, we find
\begin{align}
    H_\mr{NL} =
    & \sum_{X, X'} \int d^4 k \,
    \Bigl\{ J_{X X'}^\mr{SFWM}(k_1, k_2, k_3, k_4)
    \nonumber \\
    & \times [a_{S X k_1}^\mr{out}]^\dagger [a_{I X' k_2}^\mr{out}]^\dagger a_{P L k_3}^\mr{in} a_{P L k_4}^\mr{in} \Bigr\} + \hc,
\end{align}
where
\begin{align}
    J_{X X'}^\mr{SFWM}(k_1, k_2, k_3, k_4) \equiv
    & - \frac{3}{\epsilon_0} \int d^3x \; \Gamma_3^{lmno}
    [ D^{\mr{out}, \, l}_{S X k_1} ]^*
    \nonumber \\
    & \times
    [ D^{\mr{out}, \, m}_{I X' k_2} ]^* \,
    D^{\mr{in}, \, n}_{P L k_3} \,
    D^{\mr{in}, \, o}_{P L k_4}.
\end{align}

Decomposing the modes as in Eq.~\eqref{eq:asy-modes-ampl} and considering that the longitudinal amplitude is given by Eq.~\eqref{eq:f-def} gives us
\begin{align}
    & J_{X X'}^\mr{SFWM}(k_1, k_2, k_3, k_4) =
    \nonumber \\
    & \quad - \frac{3 \hbar^2}{16 \pi^2 \epsilon_0} \,
    \sqrt{\omega_{S k_1} \omega_{I k_2} \omega_{P k_3} \omega_{P k_4}} \,
    \Gamma_\perp^\mr{SFWM} \, I_{X X'}^\mr{SFWM},
\end{align}
where
\begin{align}
    \Gamma_\perp^\mr{SFWM} \equiv
    & \int d^2 x \,
    \Gamma_3^{lmno}(\mb{r}_\perp)
    [d_{S X k_1}^{\mr{out}, l}(\mb{r}_\perp)]^*
    \nonumber \\
    &\times
    [d_{I X' k_2}^{\mr{out}, m}(\mb{r}_\perp)]^*
    d_{P L k_3}^{\mr{in}, n}(\mb{r}_\perp)
    d_{P L k_4}^{\mr{in}, o}(\mb{r}_\perp)
\end{align}
and
\begin{align}
    I_{X X'}^\mr{SFWM} & \equiv
    \int_{-\ell/2}^{\ell/2} dz \,
    [\phi_{S X k_1}^\mr{out}(z)]^*
    [\phi_{I X' k_2}^\mr{out}(z)]^*
    \nonumber \\
    & \quad \times \phi_{P L k_3}^\mr{in}(z)
    \phi_{P L k_4}^\mr{in}(z)
    \\
    & \approx 
    \ell \, \sinc (\tfrac{\Delta K \ell}{2})
    \nonumber \\
    & \quad \times \Bigl(
    [e_{+, S X k_1}^\mr{out}]^*
    [e_{+, I X' k_2}^\mr{out}]^*
    e_{+, P L k_3}^\mr{in}
    e_{+, P L k_4}^\mr{in}
    \nonumber \\
    & \quad + [e_{-, S X k_1}^\mr{out}]^*
    [e_{-, I X' k_2}^\mr{out}]^*
    e_{-, P L k_3}^\mr{in}
    e_{-, P L k_4}^\mr{in}
    \Bigr),
\end{align}
where $\Delta K \equiv K_3 + K_4 - K_1 - K_2$.

Assuming the pump is undepleted and in the CW regime, we can follow the steps in Sec.~\ref{sec:cw-regime} to calculate the spectral generation rates for the signal photons,
\begin{align}
    &S_{X X'}^\mr{SFWM}(k_1) =
    \nonumber \\
    & \quad \frac{2 \pi \abs{\alpha_P}^4}{c \hbar^2} \abs{J_{X X'}^\mr{SFWM}(k_1, 2k_P-k_1, k_P, k_P)}^2.
    \label{eq:rate-sfwm}
\end{align}

\bibliographystyle{apsrev4-2}
\bibliography{refs}

\end{document}